\documentclass[12pt,letterpaper]{article}
{\setlength{\paperheight}{11in} \setlength{\paperwidth}{8.5in}}
\usepackage{graphicx}
\usepackage{graphics}
\begin{document}
\pagestyle{myheadings}
\title{Applications of tripled chaotic maps in cryptography}
\author{Sohrab Behnia$^{a}$\thanks{E-mail:
s.behnia@iaurmia.ac.ir}, Afshin Akhshani$^{a}$, Amir Akhavan$^{b}$, Hadi Mahmodi$^{a}$ \\
$^a${\small \emph{Department of Physics, IAU, Urmia, Iran}}\\
$^b${\small \emph{Department of Engineering, IAU, Urmia, Iran}}}
\date{}
\maketitle
\begin{abstract}
Security of information has become a major issue during the last
decades. New algorithms based on chaotic maps were suggested for
protection of different types of multimedia data, especially digital
images and videos in this period. However, many of them
fundamentally were flawed by a lack of robustness and security. For
getting higher security and higher complexity, in the current paper,
we introduce a new kind of symmetric key block cipher algorithm that
is based on \emph{tripled chaotic maps}. In this algorithm, the
utilization of two coupling parameters, as well as the increased
complexity of the cryptosystem, make a contribution to the
development of cryptosystem with higher security. In order to
increase the security of the proposed algorithm, the size of key
space and the computational complexity of the coupling parameters
should be increased as well. Both the theoretical and experimental
results state that the proposed algorithm has many capabilities such
as acceptable speed and complexity in the algorithm due to the
existence of two coupling parameter and high security. Note that the
ciphertext has a flat distribution and has the same size as the
plaintext. Therefore, it is suitable for practical use in secure
communications.
\end{abstract}
\section{Introduction}
Chaos theory is a blanketing theory that covers most aspects of
science, hence, it shows up everywhere in the world today:
mathematics, physics, biology, finance, computer and even music. As
we know, chaotic systems have many interesting features, such as the
sensitivity to the initial condition and control parameter,
ergodicity and mixing property~\cite{1,2,3,4}, which can be
connected with some cryptographic properties of good ciphers, such
as confusion/diffusion, balance and avalanche property~\cite{5,6}.
These properties make the chaotic systems a worthy choice for
constructing the cryptosystems. Compared with traditional
cryptosystems~\cite{7}, the ones based on chaos are more suitable
for large-scale data encryption such as images, videos or audio
data. Furthermore, chaos-based algorithms have shown their good
performance. Thus, it is a natural idea to use chaos as a new source
to construct new encryption systems. Cryptography has become one of
the fields of modern science since the celebrated Shannon's
work~\cite{8}. In general, the algorithms used for cryptographic
applications are classified into two, namely, public key (Asymmetric
methods) cryptography and secret key (Symmetric methods)
cryptography. According to the structure of the symmetric algorithm,
ciphers can be divided into two categories, namely, block
ciphers~\cite{9} and stream ciphers~\cite{10}. The chaotic
cryptography technique which we are going to use in this paper
belongs to a symmetric block cipher. There exist two main approaches
of designing chaos-based cryptosystems: analog and digital. The
first one is generally based on the concept of chaotic
synchronization, initiated by the work of Pecora and
Carroll~\cite{11}, and the concerned chaotic systems are implemented
in analog form. The second one is independent of chaos
synchronization and the chaotic systems are completely implemented
in digital computers~\cite{4,12}. The present paper mainly focuses
on the digital chaotic ciphers.\\ In the digital world, which is
currently evolving and changing at such a rapid pace, the security
of digital information has become increasingly more important. This
is due to the communications of digital products over network
occurring more and more frequently. So that in recent years, many
different chaotic cryptosystems in digital domain have been
proposed~\cite{13,14,15,16,17,18,19}. Certainly, a main issue in all
encryption techniques is their security.  However, the recent
development of chaotic cryptosystem is rather disappointing. On the
other hand, various cryptanalysis have exposed some inherent
drawbacks of chaotic cryptosystems~\cite{20,21,22,23}.
Some of these shortcomings are overviewed in~\cite{24,25,26}.\\
For enhancing the security of discrete chaotic cryptosystem in this
paper for the first time, we introduce the concept of using
hierarchy of tripled chaotic maps with an invariant measure in
cryptography. Six points make this new cryptosystem distinctive and
advantageous compared to the other schemes.

\begin{itemize}
  \item A very large number of fully developed chaotic
  maps defined in two intervals $x\in [0, 1]$ and $x\in [0,\infty)$.
  \item Bifurcation without any period doubling; bifurcation from a stable single periodic state to chaotic
  ones without having usual period doubling or period-n-tupling
  scenario.
  \item The existence of the two coupling parameters.
 One advantage of using two coupling parameters is for computational complexity goal. Another advantage is the
  usage of coupling parameters in the structure of the cryptosystem for diffusion. These parameters can be used as
   secret keys as well.
  \item High complexity due to high dimensionality and chaoticity.

  \item Large key space; It is obvious that the attack complexity is determined by the size of the key space and
        the complexity of the verification of each key.
   \item The flexibility of attributing different values to the
         control parameters.
\end{itemize}

Besides the above advantages, the encryption speed is also
acceptable and the ciphertext has the same size as the plaintext.
In fact, the results presented in this paper totalize our earlier work~\cite{27}.\\
This paper has been organized as follows: Section 2 describes the
construction and the main features of the tripled chaotic maps.
Current encryption schemes are introduced briefly in section 3. The
experimental results are shown in section 4. Section 5 demonstrates
the security of the proposed scheme. Finally, conclusions are drawn
in Section 6.
\section{Tripled Chaotic Maps}
We first review  the one parameter families of trigonometric chaotic
maps which are used to construct  the tripled chaotic maps.
One-parameter families of chaotic maps of the interval [0, 1] with
an invariant measure can be defined as the ratio of polynomials of
degree N~\cite{28}:
\begin{eqnarray}
\Phi_{N}^{(1,2)}(x,\alpha)=\frac{\alpha^2F}{1+(\alpha^2-1)F},
\end{eqnarray}
Where {\bf F} substitute  with chebyshev polynomial of type one
$T_{N}(x)$ for $\Phi_{N}^{(1)}(x,\alpha)$ and chebyshev polynomial
of type two $U_{N}(x)$ for $\Phi_{N}^{(2)}(x,\alpha)$\cite{29}. As
an example we give below some of these maps:
\begin{eqnarray*}
\phi_{2}^{(1)}=\frac
{\alpha^{2}(2x-1)^{2}}{4x(1-x)+\alpha^{2}(2x-1)^{2}}, \quad
\phi_{2}^{(2)}=\frac{4\alpha^{2}x(1-x)}{1+4(\alpha^{2}-1)x(1-x)},
\end{eqnarray*}
 The map $\Phi_{2}^{(2)}(x,\alpha)$ is reduced to logistic one with $ \alpha=1$.
One can show that these maps have two interesting properties. It is
shown  that these maps have interesting property, that is, for even
values of $N$ the $\Phi^{(1)}(\alpha,x)$($\Phi^{(2)}(\alpha,x)$)
maps   have only a fixed point attractor $x=1$($x=0$) provided that
their parameter belongs to interval $(N,\infty)$($(0, \frac{1}{N})$)
while, at $\alpha\geq N$ $(\alpha\geq\frac{1}{N})$ they bifurcate to
chaotic regime without having any period doubling or
period-n-tupling scenario and remain chaotic for all
$\alpha\in{(0,N)}$ ($\alpha\in(\frac{1}{N}, \infty)$) but for odd
values of N, these maps have only fixed point attractor at $x=0$ for
$\alpha\in(\frac{1}{N}, N)$, again they bifurcate to a chaotic
regime at $\alpha\geq\frac{1}{N}$, and remain chaotic for
$\alpha\in(0, \frac{1}{N})$,
 finally they bifurcate at $\alpha=N$ to have $x=1$ as fixed point
attractor for all $\alpha\in(\frac{1}{N}, \infty)$. We used their
conjugate or isomorphic maps. Conjugacy means that the invertible
map $ h(x)=\frac{1-x}{x}, $ maps $ I=[0,1]$ into $ [0,\infty) $ and
transform maps $\Phi_{N}(x,\alpha)$ into
$\tilde{\Phi}_{N}(x,\alpha)$ defined as:
\begin{eqnarray}
\tilde{\Phi}^{(1)}_{N}(x,\alpha)=\frac{1}{\alpha^{2}}\tan^{2}(N\arctan\sqrt{x}),
\label{eq:1}
\end{eqnarray}
 \begin{eqnarray}
\tilde{\Phi}^{(2)}_{N}(x,\alpha)
=\frac{1}{\alpha^{2}}\cot^{2}(N\arctan\frac{1}{\sqrt{x}}),
\label{eq:2}
  \end{eqnarray}
Using the hierarchy of families of one-parameter chaotic maps $(1)$,
we can generate new hierarchy of tripled maps with an invariant
measure. Hence, by introducing a new parameter ${\bf \epsilon}$ as a
coupling parameter we form coupling among the above mentioned maps,
where they can be coupled through $\bf{\beta}$ and ${\bf \alpha}$
functions defined as:
\begin{eqnarray}
  \beta(x(n))=(\sqrt{\beta_{0}}+\epsilon x(n))^{2},
\end{eqnarray}
\begin{eqnarray}
   \alpha_{N}(x(n))=\frac{B_{N}(\frac{1}{\beta(x(n))})}{A_{N}(\frac{1}{\beta(x(n))})}\sqrt{\frac{\beta(x(n+1))}{\beta(x(n))}},
  \end{eqnarray}
  with $B_{N}(x)$ and $A_{N}(x)$ defined as:
\begin{eqnarray}
 A_{N}(x)=\sum_{k=0}^{[ \frac{N}{2}]}C_{2k}^{N}x^{k},\quad
B_{N}(x)=\sum_{k=0}^{[ \frac{N-1}{2}]}C_{2k+1}^{N}x^{k},
\end{eqnarray}
Now, the hierarchy of the tripled chaotic maps can be defined as:
\begin{eqnarray}
\Phi_{N_{1},N_{2},N_3}=
\left\{\begin{array}{l}x_{{1}}(n+1)=\Phi_{N_{1}}(x_{1}(n),\alpha_{1}(x_{2}(n),x_{3}(n))),\\
 x_{{2}}(n+1)=\Phi_{N_{2}}(x_{2},\alpha_{2}(x_{3}(n))),\\
 x_{{3}}(n+1)=\Phi_{N_{3}}(x_{3}(n),\alpha_{3}),
\end{array}\right.
\end{eqnarray}
Also, their conjugate or isomorphic maps by considering the map
$h(x)=\frac{1-x}{x}$ are defined as follows:
\begin{eqnarray}
\tilde{\Phi}_{N_{1},N_{2},N_3}= \left\{\begin{array}{l}
\tilde{x}_{{1}}(n+1)=
\frac{1}{\alpha_{1}^2(x_{2}(n),x_{3}(n))}\tan^{2}(N_{1}\arctan\sqrt{x_{1}(n)}),\\
\tilde{x}_{{2}}(n+1)=
\frac{1}{\alpha_{2}^2(x_{3}(n))}\tan^{2}(N_{2}\arctan\sqrt{x_{2}(n)}),\\
 \tilde{x}_{{3}}(n+1)=
 \frac{1}{\alpha_{3}^{2}}\tan^{2}(N_{3}\arctan\sqrt{x_{3}(n)}),
     \end{array}\right.
 \end{eqnarray}
As we know, each dynamical system has specific characteristics. In
fact, the differences between discrete dynamical systems arise from
these properties. Due to this fact, in this paper, tripled chaotic
maps are constructed for the first time. So, it is necessary to
describe them through the paper. Specially, the most important
characteristics such as invariant measure and Lyapunov
characteristic exponents have been discussed.\\
Furthermore, many properties of the chaotic systems have their
corresponding counterparts in traditional cryptosystems, such as:
ergodicity and confusion, sensitivity to initial conditions/control
parameter and diffusion~\cite{30}. For tripled chaotic maps, we have
tried to describe ergodicity from the invariant measure point of
view.
\subsection{Invariant measure} Dynamical systems, even
apparently simple dynamical systems which are described by maps of
an interval can display a rich variety of different asymptotic
behavior. On measure theoretical level these type of behavior are
described by SRB \cite{31,32} or invariant measure describing
statistically stationary states of the system. The probability
measure ${\bf \mu}$ on $[0,1]$ is called an SRB or invariant measure
of the map $y=\Phi_{N}(x,\alpha)$ given in Eqs. (2)and(3), if it is
$\Phi_N(x,\alpha)$-invariant and absolutely continuous with respect
to Lebesgue measure. For deterministic system such as
$\Phi_{N}(x,\alpha)$-map, the $\Phi_N(x,\alpha)$-invariance means
that, its invariant measure $\mu(x)$ fulfills the following formal
FP integral equation:
$$\mu(y)=\int_{0}^{1}\delta(y-\Phi_N(x,\alpha))\mu(x)dx,$$
 This is equivalent to:
\begin{eqnarray}
\mu(y)=\sum_{x\in\Phi_{N}^{-1}(y,\alpha)}\mu(x)\frac{dx}{dy},
\end{eqnarray}
defining the action of standard FP operator for the map
$\Phi_N(x,\alpha)$ over a function as:
\begin{eqnarray}
P_{\Phi_{N}}f(y)=\sum_{x\in
\Phi_{N}^{-1}(y,\alpha)}f(x)\frac{dx}{dy},
\end{eqnarray}
We see that, the invariant measure $\mu(x)$ is actually the
eigenstate of the FP operator $P_{\Phi_N}$ corresponding to largest
eigenvalue 1. As it is proved, in our previous work, the invariant
measure $\mu_{\Phi_{N}(x,\alpha)}(x,\beta_0)$ has the following form
(for more detail see~\cite{28}):
\begin{eqnarray}
\mu{(x)}=\frac{1}{\pi}\frac{\sqrt{\beta_0}}{\sqrt{x(1-x)}(\beta_0+(1-\beta_0)x)},
\end{eqnarray}
with $\beta_0 > 0$, is the invariant measure of maps Eqs. (2)and(3)
provided that, we choose the parameter $\alpha$, in the following
form :
\begin{eqnarray}
\alpha=\frac{\sum{^{[\frac{(N-1)}{2}]}_{k=0}
C^{N}_{2k+1}\beta_0^{-k}}}{\sum{^{[\frac{N}{2}]}_{k=0}
C^{N}_{2k}\beta_0^{-k}}},
\end{eqnarray}
in $\Phi_N^{(1,2)}(x,\alpha)$  maps for odd values of N and
\begin{eqnarray}
\alpha=\frac{\beta_0\sum{^{[\frac{(N)}{2}]}_{k=0}
C^{N}_{2k}\beta_0^{-k}}}{\sum{^{[\frac{(N-1)}{2}]}_{k=0}
C^{N}_{2k+1}\beta_0^{-k}}},
\end{eqnarray}
in $\Phi_N^{(2)}(x,\alpha)$ maps for even values of N. Similarly the
probability measure ${\bf \mu}$ for tripled chaotic maps
$\Phi_{N_{1},N_{2},N_{3}}$ given in Eq. (7) fulfills the following
formal FP integral equation, where for simplicity we consider their
conjugate maps Eq. (8):

\begin{eqnarray*}
\mu(y_{1},y_{2},y_{3})= \int dx_{1}\int dx_{2}\int dx_{3}
\delta(y_{1}-\tilde{x}_{{1}}(x_{1},x_{2},x_{3}))\delta(y_{2}-\tilde{x}_{{2}}(x_{1},x_{2}))
\end{eqnarray*}
\begin{eqnarray*}
\times \delta(y_{3}-\tilde{x}_{{3}}(x_{3}))\mu(x_{1},x_{2},x_{3}),
\end{eqnarray*}
 which is equivalent to:
\begin{eqnarray}
  \mu(y_{1},y_{2},y_{3})=\sum_{x \varepsilon \tilde{\Phi}^{-1}_{N_{1},N_{2},N_{3}}}{\mid J(x_{1},x_{2},x_{2}),
  \mid}\mu(x_{1},x_{2},x_{3}),
\end{eqnarray}
 where $J(x_{1},x_{2},x_{2})$ is the Jacobian of the transformation
 $\tilde{\Phi}_{N_{1},N_{2},N_{3}}$ which is equal to:
  $$ J(x_{1},x_{2},x_{2})=\frac{\partial (\tilde{x}_{1},\tilde{x}_{2},\tilde{x}_{3})}{\partial (y_{1},y_{2},y_{3})},$$
 again defining the action of standard FP operator for the map
$\Phi_{N_{1},N_{2},N_{3}}$ over a function as:
\begin{eqnarray}
P_{\tilde{\Phi}_{N_{1},N_{2},N_{3}}}f(y)=\sum_{x\in
\tilde{\Phi}_{N_{1},N_{2},N_{3}}^{-1}}f(x)J(x_{1},x_{2},x_{3}),
\end{eqnarray}
We see that, the invariant measure $\mu(x_{1},x_{2},x_{3})$ is also
the eigenstate of the FP operator
$P_{\tilde{\Phi}_{N_{1},N_{2},N_{3}}}$ corresponding to largest
eigenvalue 1. As it is proved in Appendix A the invariant measure
$\mu_{\tilde{\Phi}_{N_{1},N_{2},N_{3}}}(x_{1},x_{2},x_{3})$ has the
following form:
\begin{eqnarray*}
\hspace{-0.59 cm}\mu=
\frac{1}{\pi}\frac{\sqrt{\beta_0}}{\sqrt{x_{3}(1-x_{3})}(\beta_0+(1-\beta_0)x_{3})}
\times\frac{1}{\pi}\frac{\sqrt{\beta(x_{3})}}{\sqrt{x_{2}(1-x_{2})}(\beta(x_{3})+(1-\beta(x_{3}))x_{2})}
\end{eqnarray*}
\begin{eqnarray}
\times
\frac{1}{\pi}\frac{\sqrt{\beta(x_{2},x_{3})}}{\sqrt{x_{1}(1-x_{1})}(\beta(x_{2},x_{3})+(1-\beta(x_{2},x_{3}))x_{1})},
\end{eqnarray}
with $\beta>0$ and  given in Eq. (4) and Eqs. (12)and(13).

\subsection{Lyapunov characteristic exponent} The method of measuring
disorder in a dynamical system, is based on the concept of Lyapunov
Characteristic Exponents (LCE). For one-dimensional maps, LCE
characterizes the local stretching, determined by the mapping
$d\phi_{n}(x,\alpha)$, weighted by the probability of encountering
that amount of stretching, that is probability of a trajectory
visiting a particular location $\mathbf{x}$. Here we compute LCE for
tripled chaotic maps as the characteristic exponent of the rate of
average magnification of the neighborhood of an arbitrary point
$\vec{r}_{0}=(x_{1_{0}},x_{2_{0}},x_{3_{0}})$ and it is denoted by
$\lambda(\overrightarrow{r_{0}})$, which can be written as
\cite{32}:
\begin{eqnarray*}
\hspace{-1 cm}\lambda(\vec{r}_{0})= \lim_{n \rightarrow
\infty}\sum_{k=0}^{n-1}
 \ln{\mid\frac{\partial ((\overbrace{\Phi \circ \Phi \circ ....\circ
\Phi}^{k})_{1} ,(\overbrace{\Phi \circ \Phi \circ ....\circ
\Phi}^{k})_{2},\overbrace{\Phi \circ \Phi \circ ....\circ
\Phi}^{k})_{3})}
 {\partial (x_{1_{0}},x_{2_{0}},x_{3_{0}})}\mid}
\end{eqnarray*}
\begin{eqnarray}
\lambda(\vec{r}_{0})=\lim_{n \rightarrow \infty}\sum_{k=0}^{n-1}
\ln{\mid \frac{\partial x_{N_{1}}(x_{k},\alpha_{1})}{\partial
x_{1_{0}}}.\frac{\partial x_{N_{2}}(x_{k},\alpha_{2}) }{\partial
x_{2_{0}}}.\frac{\partial x_{N_{3}}(x_{k},\alpha_{3}) }{\partial
x_{3_{0}}}\mid}
\end{eqnarray}
where $ x_{k}=\overbrace{\Phi \circ \Phi \circ ....\circ \Phi}^{k}$
.  A positive LCE implies that two nearby trajectories exponentially
diverge (at least locally). Negative LCE indicates contraction along
certain directions, and zero LCE indicates that along the relevant
directions there is neither expansion nor contraction. The equality
of KS-entropy and sum of all positive LCE;
$$h_{KS}=\sum_{\lambda_{l}>0}\lambda_{l} ,$$
 indicates that in chaotic region, this map is ergodic as Birkhoff
ergodic theorem predicts~\cite{33}.
\section{The Encryption and Decryption Procedures}
In this section the framework of our proposed algorithm is
described. The proposed cryptosystem is a symmetric key block cipher
algorithm based on tripled chaotic maps. Note that the Lyapunov
characteristic exponent is positive, that is, the tripled maps are
chaotic in nature. According to~\cite{34}, a possible way to
describe the key space might be in terms of positive Lyapunov
exponents. For this reason we have selected all of the control
parameters in the chaotic region. A block diagram illustrating the
complete procedure of the proposed scheme is depicted in Fig. 1.
First, plaintext is divided into blocks of the same sizes (8-bit)and
the blocks are transformed into a matrix M$_{1 \times n}$, .In
matrix M each element of the matrix represents one block, and n is
the total number of the blocks. The tripled maps are iterated using
coupling parameters, control parameters and initial conditions. In
the iteration process first 100 iterations are ignored to avoid
transient effects. The matrix element M$_{1 \times i}$ in each round
is encrypted as below:
\begin{center}
$C_{1 \times i}$ =$M_{1 \times i}$ XOR ($\tilde{x}_{f}$ mod 256),
\end{center}
where $\tilde{x}_{f}$ is a long integer generated using
$\tilde{x}_{1}$, $\tilde{x}_{2}$ and $\tilde{x}_{3}$ and some simple
mathematical operators. In each round coupling parameters and
$\tilde{x}_{f}$ are regenerated using $C_{1 \times i-1}$,
$\tilde{x}_{1}$, $\tilde{x}_{2}$ and some simple mathematical
operators. While the process reaches the last element, the elements
of matrix $C_{1 \times n}$ are reversed and $M_{1 \times n}$ is set
equal to new matrix $C_{1 \times n}$. Then again the process starts
from the beginning. In fact, the plaintext is encrypted once from
the beginning to the end and once from end to the beginning. Thus, a
very small change in the plaintext will result in a completely
different ciphertext. In this cryptosystem, the process of
decryption is completely similar to the encryption process. For
decryption process, the only difference is that the following new
relation is used instead of aforesaid relation.
\begin{center}
$M_{1 \times i}$=$C_{1 \times i}$ XOR ($\tilde{x}_{f}$ mod256),
\end{center}
As an example, the following maps are selected from the chaotic
maps(Eq.8) to be used in encryption/decryption process.
\begin{eqnarray}
\tilde{\Phi}_{2,2,14}(\tilde{x}_{1},\tilde{x}_{2},\tilde{x}_{3})=
\left\{
\begin{array}{l} \tilde{x}_{1}(n+1)=
\frac{1}{\alpha_{1}^2(x_{2}(n),x_{3}(n))}\tan^{2}(2\arctan\sqrt{x_{1}(n)}),\\
\tilde{x}_{2}(n+1)=
\frac{1}{\alpha_{2}^2(x_{3}(n))}\tan^{2}(2\arctan\sqrt{x_{2}(n)}),\\
 \tilde{x}_{3}(n+1)=
 \frac{1}{\alpha_{3}^{2}}\tan^{2}(14\arctan\sqrt{x_{3}(n)}),
     \end{array}\right.
\end{eqnarray} with
$$
\alpha_{2}(x_{3}(n))=\frac{2\beta(x_{3}(n))}{1+\beta(x_{3}(n))}\sqrt{\frac{\beta(x_{3}(n+1))}
{\beta(x_{3}(n))}}, \quad
\beta(x_3(n))=(\sqrt{\frac{\alpha_{2}}{2-\alpha_{2}}}+ \epsilon
x_3(n))^{2},
$$
$$
\alpha_{1}(x_{2}(n),x_{3}(n))=\frac{2\beta(x_{2}(n),x_{3}(n))}{1+\beta(x_{2}(n),x_{3}(n))}\sqrt{\frac{\beta(x_{2}(n+1),x_{3}(n+1)))}
{\beta(x_{2}(n),x_{3}(n))}},
$$
$$ \beta(x_{2}(n),x_{3}(n))=(\sqrt{\frac{\alpha_{1}}{2-\alpha_{1}}}+ \epsilon'
x_2(n))^{2},
$$
\section{Experimental Results}
In this section, we provide some experimental results to illustrate
the performance of the proposed chaotic cryptosystem. In order to
test the efficiency of the proposed chaotic cryptographic scheme, we
used the scheme in the following files.
\begin{description}
  \item File 1: Text (.txt) file of size 30 720 bytes;
  \item File 2: Word document (.doc) file of size  210 944 bytes;
  \item File 3: Executable (.exe) file of size  487 000 bytes;
  \item File 4: Audio (.mp3) file of size 980 304 bytes;
  \item File 5: Image (.bmp) file of size 65 536 bytes;
  \item File 6: Video clip (.avi) file of size 1 087 430 bytes.
\end{description}
The encryption and decryption time for all of source files mentioned
above are listed in Table 1.\\
The length of the ciphertext is the same as thought the plain text.
This features is another most important features of our introduced
cryptosystem(See the last column of Table 1). Note that, the
previous cryptosystems like Baptista-type chaotic cryptosystems
\cite{35,36,37} had almost double-sized ciphertext.\\ In order to
compare the performance evaluation of the proposed method with the
previous work~\cite{27}, from the security point of view, we focus
on the application of this method in image encryption. We assume
that source image is 65 536 bytes. Fig. 2(a) shows the experimental
results with Bird BMP image. Fig. 2(b) is its encrypted image with
the encryption keys mentioned below. According to Eq. (18),
encryption keys are chosen as follows: $\tilde{x}_{3}$=66,
$\tilde{x}_{2}$=444, $\tilde{x}_{1}$=445 as initial conditions and
$\epsilon$=0.2, $\epsilon'$=0.5 as a coupling parameters and finally
$\alpha$$_{3}$=1.5, $\alpha$$_{2}$=0.455, $\alpha$$_{1}$=0.2 as
control parameters. We have implemented the proposed algorithm using
C++ programming language and observed the simulation results on a
Pentium-IV 2.4 GHz Celeron D with 256MB RAM and 80 Gb hard-disk
capacities. It seems that, our encryption time is  acceptable
compared to that of encryption times mentioned in~\cite{38,39,40}.
\section{Security Analysis}
When a new cryptosystem is proposed, it should always be accompanied
by some security analysis. A good encryption procedure should be
robust against all kinds of cryptanalytic, statistical and
brute-force attacks. Here, some security analysis has been performed
on the proposed scheme, including some important ones like key space
analysis, statistical analysis, etc. The security analysis
demonstrated the high security of the new scheme, as demonstrated in
the following.
\subsection{Key space analysis}
A fundamental aspect of every cryptosystem is the key. An algorithm
is as secure as its key. No matter how strong and well designed the
algorithm might be, if the key is poorly chosen or the key space is
small enough, the cryptosystem will be broken. The size of the key
space is the number of encryption/decryption key pairs that are
available in the cipher system. Apparently, the attack complexity is
determined by the size of the key space and the complexity of
verifying each key. From the cryptographical point of view, the size
of the key space should not be smaller than $2^{100}$ to provide a
high level of security~\cite{6,41}. If the precision 10$^{-16}$, the
key space size for initial conditions, control parameters and
coupling parameters is over than $2^{400}$. It seems that the key
space is large enough to resist all kinds of brute-force attacks.\\
In addition, one attempt to describe the dynamics of the current
system is by providing bifurcation diagram and determining
interesting properties of it. In any case, the designer of any
chaotic cryptosystem should conduct a study of chaotic regions of
the parameter space from which valid keys, i.e., parameter values
leading to chaotic behavior, can be chosen. When many parameters are
used simultaneously as part of the key, the mutual interdependence
complicates the task of deciding which intervals are suitable. Only
keys chosen from the black region of bifurcation diagram are
suitable enough~\cite{42,43}. In this cryptosystem, the interval of
the initial condition is $[0,\infty)$. From the bifurcation diagram
it is clear that the maps are chaotic for any $x$ with respect to
control parameters in chaotic region. In Fig. 3, we have depicted a
portion of the bifurcation diagram of
$\tilde{\Phi}^{(1)}_{N}(x,\alpha)$ while $N$=2. As we can see,
within the black region, there isn't any periodic windows, so the
entire black region is suitable for robust keys.
\subsection{Statistical analysis}
In his masterpiece, Shannon~\cite{8} said, ``It is possible to solve
many kinds of ciphers by statistical analysis," and therefore, he
suggested two methods of diffusion and confusion for the purpose of
frustrating the powerful statistical analysis. This is shown by a
test on the histograms of the ciphered images, on the correlations
of adjacent pixels in the ciphered image and on the distribution of
the ciphertext.\\ \emph{1.Histograms of ciphered images.} By taking
a (256$\times$256) sized Bird image as a plaintext, the histograms
of the plaintext and its corresponding ciphertext are as follows. It
is clear that the encrypted image is confused and cannot be
understood. Moreover, the histogram of the encrypted image is
uniformly distributed, which makes statistical attacks very
difficult (See Figs. 4(a) and 4(b)).\\ \emph{2. Correlation of two
adjacent pixels.} Statistical analysis has been performed on the
proposed image encryption algorithm by a test on the correlation of
adjacent pixels in the plain-image and ciphered image. To analyze
the correlations of the adjacent pixels, we can use Eq. (19) to
calculate the correlation coefficients in horizontal, vertical and
diagonal~\cite{14,44}.
\begin{eqnarray}
cov(x,y)=\frac{1}{N}\sum_{i=1}^{N} (x_{i}-E(x))(y_{i}-E(y)), \quad
r_{xy}=\frac{cov(x,y)}{\sqrt{D(x)}\sqrt{D(y)}}, \label{eq:14}
\end{eqnarray}
where
$$
E(x)=\frac{1}{N}\sum_{i=1}^{N} x_{i}, \quad \quad \quad
D(x)=\frac{1}{N}\sum_{i=1}^{N} (x_{i}-E(x))^2. \label{eq:12}
$$
$E(x)$ is the estimation of mathematical expectations of $x$, $D(x)$
is the estimation of variance of $x$ and $cov(x,y)$ is the
estimation of covariance between $x$ and $y$. where $x$ and $y$ are
grey-scale values of two adjacent pixels in the image. We randomly
choose 1000 image pixels in the plain image and the ciphered image
respectively to calculate the correlation coefficients of the
adjacent pixels in horizontal. The correlation coefficients of the
adjacent pixels in vertical and in diagonal are calculated and
listed in Table 2 and the distribution is shown in Figs. 5(a) and
5(b).
It demonstrates that the encryption algorithm has covered up all the
characters of the plain image and shows good performance of balanced
0-1 ratio.\\In addition to analysis mentioned above, we have also
analyzed the distribution of the ciphertext. Therefore, we have
recorded the number of occurrences of each ciphertext block for one
of the six source files. A typical distribution of the ciphertext is
shown in Figs. 6(a) and 6(b), which shows that the distribution is
very flat due to the masking operation. Totally, statistical
analysis has been performed on the proposed encryption algorithm,
demonstrating its superior confusion and diffusion properties which
strongly resist statistical attacks.
\subsection{Information entropy}
Information theory is a mathematical theory of data communication
and storage founded in 1949 by Claude E. Shannon. To calculate the
entropy \textit{H(s)} of a source \textit{s}, we have:
\begin{eqnarray}
H(s)=\sum_{i=0}^{2N-1}P(s_{i})\log_{2} \frac{1}{P(s_{i})},
\label{eq:10}
\end{eqnarray}
where $P(s_{i})$ represents the probability of symbol $s_{i}$.
Actually, given that a real information source seldom transmits
random messages, in general, the entropy value of the source is
smaller than the ideal one. However, when these messages are
encrypted, their entropy should ideally be 8. If the output of such
a cipher emits symbols with an entropy of less than 8, then there
exists a predictability which threatens its security. We have
calculated the information entropy for encrypted image Fig. 2(b):
\[
H(s)=\sum_{i=0}^{255}P(s_{i})\log_{2}\frac{1}{P(s_{i})}=7.9978
\]
The obtained value is very close to the theoretical value 8.
Apparently, comparing it with the other existing algorithms, such as
~\cite{15}, the proposed algorithm is much more closer to the ideal
situation. This means that information leakage in the encryption
process is negligible, and so the encryption system is secure upon
the entropy attack.
\subsection{Differential attack}
As we know, if one minor change in the plain-image can cause a
significant change in the ciphered-image, with respect to both
diffusion and confusion, then this "differential attack" may become
inefficient. To test the influence of one-pixel change on the whole
image, encrypted by the proposed chaos-based algorithm, two common
measures may be used: Number of Pixels Change Rate (NPCR) and
Unified Average Changing Intensity (UACI)~\cite{45,46}. We take two
encrypted images, $C_{1}$ and $C_{2}$, whose corresponding original
images have only one-pixel difference. We label the grey scale
values of the pixels at grid (i,j) of $C_{1}$ and $C_{2}$ by
$C_{1}(i,j)$ and $C_{2}(i,j)$, and $C_{1}$ and $C_{2}$ have the same
size. Then, $D(i,j)$ is determined by $C_{1}(i,j)$ and C$_{2}(i,j)$,
that is, if $C_{1}(i,j)= C_{2}(i,j)$, then, $D(i,j)=1$; otherwise,$
D(i,j)=0$. NPCR and UACI are defined by the following formulas:
\begin{eqnarray*}
NPCR=\frac{\sum_{i,j}D(i,j)}{W\times H} \times 100\%
\end{eqnarray*}
\begin{eqnarray*}
UACI=\frac{1}{W \times H}\left[\sum_{i,j}\frac{|C_{1}(i,j) -
C_{2}(i,j)|}{255}\right] \times 100\%
\end{eqnarray*}
Where, $W$ and $H$ are the width and length of the image. We
obtained NPCR=0.34 \% and UACI=0.33 \%. With regard to obtained
results, apparently, the proposed algorithm has a good ability to
resist differential attack.
\section{Conclusion}
In this paper, after introducing hierarchy of tripled chaotic maps
with invariant measure, we investigated their potential for
exploitation as a cryptosystem. These maps have interesting features
such as invariant measure, ergodicity and variable control
parameters. In addition to some features aforesaid, the most
important advantage of these maps is the existence of the two
coupling parameters. It seems that the excellent efficiency of the
new cryptosystem is derived from this property. In this scheme, its
structural parameters and initial values can all be used as
encryption key in chaotic cryptosystem. Therefore, the size of key
space is very large. Statistical analysis performed on the proposed
encryption algorithm, demonstrates its superior confusion and
diffusion properties which strongly resist statistical attacks.
Experimental results illustrate that the distribution of the
ciphertext is very flat and the entropy, is almost equal to the
ideal value. Moreover, the encryption time is acceptable while the
size of ciphertext is the same as that of the plaintext. The present
paper's goal is to construct a practical and fully secure
cryptosystem. Our experiments indicate that this goal is almost
achieved. Based on all analysis and experimental results, the
conclusion is that, from a cryptographical point of view, the
proposed scheme is a good candidates for practical applications in
information security fields.
\section*{Acknowledgment}
The authors would like to express their heartfelt gratitude to Mr.
A. Bonyadi for the nice editing of their paper - this has certainly
improved its readability.
\newpage
 \renewcommand{\thesection}{A}
\renewcommand{\theequation}{\thesection.\arabic{equation}}
\setcounter{equation}{0}
 {\large \appendix{\bf Appendix A}}: {Derivation of the invariant
measure} In order to prove that measure Eq. (16) satisfies Eq. (14),
it is rather convenient to consider the conjugate map. By inverting
the conjugate map Eq. (8) we get:
\begin{equation}  \left\{
\begin{array}{l}x_{k3}=\tilde{\Phi}_{N_{3}}^{-1}(x_{3},\alpha_3)),\\
 x_{k2, k3}=\tilde{\Phi}_{N_{2}}^{-1}(x_{2},\alpha_2(x_{ k3})),\\
 x_{k1, k2, k3}=\tilde{\Phi}_{N_{1}}^{-1}(x_{1},\alpha_1(x_{k2,
 k3},x_{k3})),
\end{array}\right.
\end{equation}
Now, considering the following anzatz for invariant measure
\begin{equation}
\mu(x_1,x_2,x_3)=\frac{1}{\pi}\frac{\gamma(x_2,x_3)}{(1+\beta(x_2,x_3)x_1)\sqrt{x_1}},
\end{equation}
where
$$\int \frac{\gamma(x_2,x_3)}{\sqrt{\beta(x_2,x_3)}}dx_2dx_3=1.$$
The probability measure ${\bf \mu}$ for coupled chaotic maps
$\Phi_{N_{1},N_{2},N_{3}}$ given in Eq. (7) fulfills the following
formal FP integral equation (Eq. 14):
\begin{equation}
\hspace{-0.75 cm}\mu(x_1,x_2,x_3)=\sum_{k1,k2,k3}\mid \frac{\partial
x_{k1,k2,k3}}{\partial x_1}\frac{\partial x_{k2,k3}}{\partial
x_2}\frac{\partial x_{k3}}{\partial x_3} \mid
\mu(x_{k1,k2,k3},x_{k2,k3},x_{k3})
\end{equation}
by taking the derivative of last term of Eq. (A.1) with respect to
$x_1$, we have:
\begin{equation}
\frac{\partial x_{k1,k2,k3}}{\partial
x_1}=\frac{\alpha_1(x_{k2,k3},x_{k3})}{N_1}\frac{\sqrt{x_{k1,k2,k3}}(1+x_{k1,k2,k3})}{\sqrt{x_1}(1+\alpha_1^2(x_{k2,k3},x_{k3})x_1)}
\end{equation}
from Eq. (A.2) it follows that:
$$
\frac{\gamma(x_2,x_3)}{(1+\beta(x_2,x_3)x_1)\sqrt{x_1}}=\sum_{k1,k2,k3}\frac{\alpha_1}{N_1}
\frac{(1+x_{k1,k2,k3})}{\sqrt{x_1}(1+\alpha_1^2(x_{k2,k3},x_{k3})x_1)}
$$
\begin{equation}
\times
\frac{\gamma(x_{k2,k3},x_{k3})}{1+\beta(x_{k2,k3},x_{k3})x_{k1,k2,k3}}
\end{equation}
we readily see that
$$
\sum_{k3}\frac{\alpha_1}{N_1}
\frac{(1+x_{k1,k2,k3})}{(1+\beta(x_{k1,k2,k3}))}=\frac{\alpha_1A_{N_1}(\beta^{-1}(x_{k2,k3},x_{k3}))}
{B_{N_1}(\beta^{-1}(x_{k2,k3},x_{k3}))}
$$
\begin{equation}
\times
\frac{(1+\alpha_1^2(x_{k2,k3},x_{k3}))x_1)}{1+\beta(x_{k2,k3},x_{k3})(\frac{A_{N_1}(\beta^{-1})\alpha_3}
{B_{N_1}(\beta^{-1})})^2x_1}
\end{equation}
Eq. (A.3) reduce to:
$$\frac{\gamma(x_2,x_3)}{(1+\beta(x_2,x_3)x_1)}=\sum_{k1,k2,k3}\frac{\partial
x_{k2,k3}}{\partial x_2}\frac{\partial x_{k3}}{\partial x_3}$$
\begin{equation}
\times
\frac{\alpha_1(x_{k2,k3},x_{k3})A_{N_1}(\beta^{-1}(x_{k2,k3},x_{k3}))}
{B_{N_1}(\beta^{-1}(x_{k2,k3},x_{k3}))}\frac{1}{1+\beta(x_{k1})(\frac{A_{N_1}(\beta^{-1})\alpha_1}
{B_{N_1}(\beta^{-1})})^2x_1},
\end{equation}
which is possible if $\beta(x_2,x_3)$ and $\beta(x_{k2,k3},x_{k3})$
are related as:
\begin{equation}
\left(\frac{\alpha_1A_{N_1}(\beta^{-1}(x_{k2,k3},x_{k3}))}
{B_{N_1}(\beta^{-1}(x_{k2,k3},x_{k3}))}\right)^2\beta(x_{k2,k3},x_{k3})=\beta(x_2,x_3)
\end{equation}
therefore, the relation Eq. (A.7) reduces to
\begin{equation}
\frac{\gamma(x_2,x_3)}{\sqrt{\beta(x_2,x_3)}}=\sum_{k2,k3}\frac{\partial
x_{k2,k3}}{\partial x_2}\frac{\partial x_{k3}}{\partial
x_3}\frac{\gamma(x_{k2,k3},x_{k3})}{\sqrt{\beta(x_{k2,k3},x_{k3})x_{k3}}}=\mu(x_2,x_3)
\end{equation}
By considering
\begin{equation}
\frac{\gamma(x_2,x_3)}{\sqrt{\beta(x_2,x_3)}}=\frac{1}{\pi}\frac{\gamma(x_3)}{\sqrt{x_2}(1+\beta(x_3)x_2)}
\end{equation}
where
$$\int \frac{\gamma(x_3)}{\sqrt{\beta(x_3)}}dx_3=1$$
which is possible if $\beta(x_3)$ and $\beta(x_{k3})$ are related
as:
\begin{equation}
\left(\frac{\alpha_2(x_3)A_{N_2}(\beta^{-1}(x_{k3}))}
{B_{N_2}(\beta^{-1}(x_{k3}))}\right)^2\beta(x_{k3})=\beta(x_3)
\end{equation}
or Eq. (A.9) can be written as:
$$ \frac{\gamma(x_3)}{\sqrt{\beta(x_3)}}=\sum_{k3}\frac{\partial
x_{k3}}{\partial
x_3}\frac{\gamma(x_{k3})}{\sqrt{\beta(x_{k3})}}=\frac{1}{\pi}\frac{\beta}{\sqrt{x_3}(1+\beta
x_3)}$$ hence it is proportional to its invariant measure.

\clearpage

\begin{table}
\caption{\label{tab:1}Performance of the proposed chaotic
cryptographic on the six files, encryption/decryption time}
\begin{tabular}{cccc} \hline
    & Plaintext file size &  Encryption time (in seconds) & Ciphertext file size\\
  &      (bytes)       & min.-max. (mean) & (bytes)\\  \hline
File 1 (.txt) &  30 720   &  0.14-0.15 (0.145)  & 30 720 \\
File 2 (.doc) &  210 944   & 1.005-1.007 (1.006)   & 210 944 \\
File 3 (.exe) &  487 000  & 2.5-2.56 (2.53)   & 487 000 \\
File 4 (.mp3) &  980 304   & 0.44-0.47 (0.455)  & 980 304 \\
File 5 (.bmp) &  65 536   &  0.33-0.36 (0.345) & 65 536 \\
File 6 (.avi) &  1 087 430 & 5.61-5.67 (5.64)  & 1 087 430 \\
\hline
\end{tabular}
\end{table}
%
 \clearpage
%
\begin{table}
\begin{center}
\caption{\label{tab:2}Correlation coefficients of two adjacent
pixels in two images}
\begin{tabular}{llllll} \hline
 &  &Plain image& &  Ciphered image \\
\hline
Horizontal  && 0.9902 & & 0.000029549\\
Vertical    && 0.9815  & & 0.00045066\\
Diagonal    && 0.9721  & & 0.000013\\ \hline \noalign{\smallskip}
\end{tabular}
\end{center}
\end{table}

 \clearpage


Figures Caption:

Fig. 1: Block Diagram.

Fig. 2: (a) Plain image, (b) Ciphered image.

Fig. 3: Bifurcation diagram of $\tilde{\Phi}^{(1)}_{N}(x,\alpha)$
while $N$=2.

Fig. 4: (a) Histogram of plain image, (b) Histogram of ciphered
image.

Fig. 5: (a) Correlation analysis of plain image, (b) Correlation
analysis of ciphered image.

Fig. 6: (a) Distribution of the plaintext, (b) Distribution of the
ciphertext.

\end{document}